# A multi-criteria service selection algorithm for business process requirements

Sophea Chhun[1], Chantal Cherifi, Néjib Moalla, Yacine Ouzrout

*Lab. DISP, University of Lumiere Lyon2, Lyon, France*
*IUT Lumière – DISP Laboratory, 160 Bd de l'Université, 69676 Bron Cedex*
*firstname.lastname@univ-lyon2.fr*

**ABSTRACT**

The selection of the most appropriate Web services to realize business tasks still remain an open issue. We propose a multi-criteria algorithm for efficient service selection. Web services and their QoS values are stored in a Web service ontology (WSOnto) and business processes are modeled with the BPMN2.0 specifications. Our algorithm performs an instance-based ontology matching between the WSOnto and the business process ontology. The business context, functional properties and QoS values of Web services are considered. The algorithm computes the variation of QoS values over times. This strategy allows better accurate Web services ranking relevant to a user's request.

## 1. Introduction

A business process is an activity or a set of activities that accomplish a specific organizational goal. Activities or tasks are linked together and are executed following a pre-defined order. Business tasks can be manual tasks, service tasks, script tasks and even sub processes. A business process can realize a business goal of a single organization or a group of organizations. Legacy systems that implement enterprise business processes reach some limitations regarding nowadays ongoing business changes. Enterprises evolve in a highly dynamic environment with fast mutations. Depending on their size, their complexity, the number of partners involved, changing needs, these systems may rapidly become difficult to maintain, inefficient, and consequently costly. In order to stay competitive, enterprises need solutions designed to give the agility to cope with ongoing changes. Newer technologies and more efficient methods to implement the business tasks of a business process are required.

The modeling of business processes is an important step in the improvement of business performances. It aims at proposing an unambiguous representation of complex enterprise processes in order to optimize the efficiency of connecting activities in the provision of products or services. Different business process modeling notations and languages exist such as Business Process Modeling Notation (BPMN) [1], Petri Net [2], Workflow [3], Unified Modeling Language (UML) [4] and Business Process Modeling Language (BPML) [5]. Each language provides different notions, syntax, and complexity for modeling a business process.

Web services have been an opportunity taken by many organizations in order to modernize and to transform their legacy systems. This technology provides a mean to migrate and to externalize core business logic and competencies. A Web service, as defined by the W3C, "is a software system designed to support interoperable machine-to-machine interaction over a network". Functions available throughout an organization are presented through standard-based services within a Service Oriented Architecture (SOA) [6]. They can be incorporated in any internal or external applications. In this work, we combine the expressive power of business process modeling with the agility of Web services in a unified and valuable framework called BPMNSemAuto. BPMN is in charge of expressing business processes required by a user, while Web services selected from a registry are used to implement its various tasks.

The management of business processes tends to be even more valuable if it can be automated, and if processes are fuelled by the most accurate and up-to-date available services. To cope with this challenge, we developed an ontology based architecture for dynamic service selection. The designed business processes of users as well as the existing Web services are described by ontologies. An instance-based ontology matching between the business process ontology (BPOnto) and the Web service ontology (WSOnto) allows to find appropriate Web services to implement the desired service tasks of business processes.

This service selection step is a crucial activity within SOAs. Candidate services with similar functionalities are numerous. They are developed by different providers using their own vocabularies to name the services, their operations and parameters.

---

[1]Corresponding author. Email: sophea.chhun@univ-lyon2.fr; Tel: +337 81 85 33 16

The challenge is to select the most satisfactory services that will implement business tasks. Besides functional properties of Web services, Quality of Service (QoS) attributes are criteria for a service selection process. It is an important non-functional aspect used to differentiate functionally similar Web services. QoS embodies response time, execution time, availability, reliability, security, number of calls, etc. Some QoS properties can change over times. For example, from one call to another of a Web service, execution time may increase or decrease for various reasons. This brings a dynamic aspect important to be considered. Web service selection is a research issue that has triggered a large body of work since the early 2000s [7]. Selection algorithms take into account one or a combination among the business context (keywords expressing a category and functionalities of Web services), functional and non-functional (QoS) properties of the service. Great efforts have been made to address the problem of QoS-based service selection [8], [9]. However, most of these works fail to integrate the dynamic nature of QoS. To take into account the dynamic nature of the Internet, average values of the QoS attributes have been used [10], [11]. More recently, in order to increase the accuracy, QoS attributes are described by a range value rather than a single one. Indeed, recording the upper and lower bound of fluctuant QoS attribute allows a finer description [12]. Our work is in this line. We propose to consider the changing value of QoS over time by incorporating the variations during a time slot. The proposed selection algorithm is based on three criteria: the context, the functional properties, and the dynamic QoS attributes. Indeed, with the rapid growing number of Web services, it is difficult to find the most appropriate candidate regarding to functional properties and QoS constraints in an exhaustive search. In order to reduce substantially the cost of a selection process, the functionality score computation is restricted to Web services whose context matches a user's requirements. Functionality scores are expressed according to the similarity between input and output parameters of business tasks and Web services. Similarly, QoS scores are computed only for Web services that have functionality score at least equal to the defined minimum acceptable functional score. Finally, selected Web services are ranked according to a linear combination of functionality and QoS scores tuned by users. The algorithm follows a local optimization-based approach. Indeed, the best service for individual tasks is chosen, one task at a time, regardless of the task dependencies in a business process, or the end-to-end quality requirements of the composite service.

The main contributions of this work are: (i) the design of a service selection algorithm, including three axes: context matching, functional matching and QoS matching. The algorithm also considers the preferences QoS values of users by allowing them define weights; (ii) the introduction of the QoS dynamics in the service selection algorithm; (iii) an instance-based ontology matching for service selection.

The rest of this paper is arranged as follows. Section 2 recalls fundamental information about Business Process Modeling, Web service description and a presents a summary about existing service selection methods. Section 3 provides an overview of the instance-based matching. Section 4 is dedicated to the presentation of business process and Web service ontologies. Section 5 is devoted to a thorough presentation of the proposed service selection algorithm. Section 6 presents the implementation of the proposed framework. Section 7 concludes the paper and highlights orientations for future works.

## 2. Background

### 2.1. Business process modeling

Business process modeling is based on two dominant formalisms, graph-based and rule-based [13]. Rule-based is grounded by the formal logic. The graph-based approach provides a graphical interface enabling users to model business processes in an intuitive way. This is undoubtedly an asset which makes it far more popular than the rule-based approach.

In [13], the authors conduct a comparison study based on different criteria between several graph-based and rule-based business process modeling. Among the graphical modeling, BPMN seems to be the one that provides the greatest ease of use for business users. BPMN was presented as a standard in 2004 by the Object Management Group. It bridges the gap between the design and implementation of business processes. The primary goal of BPMN is "to provide a notation that is readily understandable by all business users, from the business analysts that create the initial drafts of the processes, to the technical developers responsible for implementing the technology that will perform those processes, and finally, to the business people who will manage and monitor those processes" [1]. The BPMN specifications divide business process elements into four categories: (i) Flow objects that define the behaviours of a business process. A flow object is an event, an activity or a gateway; (ii) Connecting objects that connect two flow objects or a flow object with other resources. There are three types of connecting objects: sequence flow, message flow and association; (iii) Swim lanes that group the primary modeling elements. There are two kinds of swim lanes, pool and lane; (iv) Artifacts allow to provide additional information about business processes. They are categorized into three sub-groups: processed data, groups of activities and annotations. Two or many business tasks are linked to each other by gateways (parallel, inclusive, exclusive, complex and event based gateway). The version 2 of BPMN was released in 2011. It has around 100 different modeling constructs, including 51 event types, 8 gateway types, 7 data types, 4 types of activities, 6 activity markers, 7 task types, 4 flow types, pools, lanes, etc. [14].

*2.2. Web service description*

In this research, Web service description is based on the Web Service Description Language (WSDL) standard [15]. WSDL specifies functional and non-functional properties of Web services. The functional properties describe operations that a service exposes with their input and output parameters. Parameters are described by their name and data type. Non-functional properties of a WSDL file concern the location of the service, the communication protocol and the data format specifications.

Other solutions have been proposed to enrich the description with semantics. Indeed, WSDL provides only syntactical information. It lacks the semantic expressiveness needed to represent Web services capabilities. This situation can lead to possible misinterpretation. The semantics are introduced by ontologies that support shared vocabularies and allow automatic reasoning. The semantic Web service field includes substantial bodies of work through three conceptual approaches, Ontology Web Language for Services (OWL-S) [16], Web Services Modeling Ontology (WSMO) [17], Semantic Annotation for WSDL (SAWSDL) [18] and WSDL-Semantic (WSDL-S) [19]. Table 1 provides an overview of Web service description languages. The criteria for comparing the languages are: (i) the approach used to introduce semantics in the description; (ii) the level of semantics; (iii) the representation of semantics; (iv) the versions; (v) the functional content. The semantics can be introduced by two different ways. Either a native language is proposed or the existing WSDL is enriched by semantics.

**Table 1**
Web service description languages.

| Web Service Description Languages | Semantic Description Approach | Semantic | Representation | Version/Year | Functional Content |
|---|---|---|---|---|---|
| **WSDL** (Web Service Description Language) | Syntactic description language | No | XML | V 1.0, 2000 | Input |
| | | | | V 1.1, 2001 | |
| | | | | V 1.2, 2003 | Output |
| | | | | V 2.0, 2007 (W3C recommendation) | |
| **WSDL-S** (Web Service Description Language-Semantic) | Annotation of existing languages | Yes (High) (Requires the domain ontology) | WSDL/XML | 2005 (W3C Member Submission) | Input |
| | | | | | Output |
| | | | | | Pre-condition |
| | | | | | Post-condition |
| **DAML-S** (DARPA Agent Markup Language-Semantic) | Semantic description language | Yes (High) | Ontology | 2003 (DAML-S Working Group) | Input |
| | | | | | Output |
| **OWL-S** (Ontology Web Language for Web Services) | Semantic description language | Yes (High) | Ontology | V 1, 2004 (W3C member submission) | Input |
| | | | | | Output |
| | | | | | Pre-condition |
| | | | | V 1. 2, 2006 | Post-condition |
| **WSMO** (Web Services Modelling Ontology) | Semantic description language | Yes (High) (Requires the domain ontology) | Ontology | 2005 (W3C Member Submission) | Input |
| | | | | | Output |
| | | | | | Pre-condition |
| | | | | | Post-condition |
| | | | | | Goal |
| **SAWSDL** (Semantic Annotations for WSDL) | Annotation of existing languages | Yes (High) (Requires the domain ontology) | WSDL/XML/RDF (Ontology) | 2007 (SAWSDL Working Group) | Input |
| | | | | | Output |

For the former case, DAML-S, OWL-S and WSMO adopt a top-down perspective. The semantics are described by using domain ontologies that are independent from the service development. These languages are grounded with the corresponding WSDL description. For the latter, WSDL-S and SAWSDL adopt a bottom-up approach where the WSDL file is annotated using semantic information. For the annotation approach, the functional content, i.e. input and output parameters, are described by domain ontologies. In the other approach, an ontology of Web service is designed. Each Web service description language allows expressing different functional content of Web services, as shown in Table 1. For example, WSDL allows expressing only input and output parameters of services operations. OWL-S is designed for expressing inputs, outputs, pre-condition and post-condition of services' operations.

Some proposals can be found in the literature to add QoS attributes to Web service descriptions. In [20], the authors extend the OWL-S profile ontology to represent QoS properties and business offer properties. The QoS properties contain business QoS properties (price, compensation, withdraw period and penalty rate), Performance QoS properties (response time, throughput, availability and security) and response QoS properties (success rate, reputation and compliance). The business offer properties are divided into unconditional business offer, conditional business offer and probabilistic business offer. In [21], the author proposes a light weight WSDL extension for the description of QoS characteristics, such as performance, reliability, availability and security.

In this work, Web services are stored in a Web service ontology called WSOnto. Its design is based on syntactic descriptions of Web services, i.e. WSDL. Ontologies have many avantages. First, it has been recognized as knowledge based. Second, it supports sharing and interoperability between partners. Third, it can be extended and modified through manual modification or by using some ontology matching and merging techniques [22], [23]. Last, it can be used to store both functional and non-functional properties of Web services.

*2.3. Service selection algorithms*

Service selection is the process of discovering and selecting the most appropriate Web services in order to respond to a user's requirements.

The traditional Web service discovery approach is a keyword-based search using the UDDI (Universal Description, Discovery and Integration) registry. Some of service discovery approaches are syntax based while others are semantic based [7]. The keyword matching service discovery is based on syntactic while ontology service discovery is semantic-based. During the discovery and selection processes, considering multiple criteria in algorithms can improve results. QoS values of Web services are used to select the best Web service out of a set of Web services that provide the same functionalities. In [24], the authors state that "neglecting QoS will cause serious problems in software development". In [25], Xianglan et al., 2011 provide a survey on QoS-based discovery and selection. For the service selection based on context, they are described in [26]. The context or goals of users are expressed in keywords. The context based service selection process chooses Web services that answer to users' goals.

Since the UDDI registry supports only keywords matching, Sycara et al., 2003 [27] extend the capabilities of the UDDI by introducing a MatchMaker. This MatchMaker can perform the semantic matching based on ontologies. Usually, an ontology match process returns four degrees of matches: exact, plugin, subsume and fail. Broens et al., 2004 [28] also use ontology to perform services ranking. Their algorithm performs four steps ontology matching in sequence to rake WSs: service type, inputs, outputs and contextual information. These criteria are all expressed in keywords. Pakari et al., 2014 [29] integrate WordNet in addition to domain ontologies. To compare between a user's requirements and information of a Web service, they use a hybrid semantic matching method that combines three comparisons: syntactic matching by using Jaro-Winkler strategy, structural matching by using WordNet taxonomy with WuAndPalmer algorithm, and semantic matching based on ontology. However, the authors of [27], [28] and [29] do not consider QoS properties in their proposed algorithms.

In [30], [31] and [32], authors focus on proposing algorithms to rank WSs that provide the same functionalities, the functional matching part was not considered. In addition, QoS properties of Web services are considered in their algorithms. Tran et al., 2009 [30] propose a very flexible QoS ontology to store QoS properties with multi level and fine-grained service level. They use an AHP method to calculate QoS scores for ranking Web services. The complexity of their algorithm depends on domain applications with QoS property specifications. Their proposed QoS ontology structure is flexible, but time consuming for two reasons. First, different QoS calculation methods are needed for difference find-grained of QoS parameters. Second, the complexity of the algorithm depends on AHP hierarchy. Cabrera et al., 2011 [31] integrate the property mandatory option to identify QoS attributes in addition to their value and minimization/maximization property (to mean users need big or small values of QoS). If it is a mandatory attribute defined, it means that the selection process must verify this constraint before selecting a Web service. These three characteristics of QoS attributes allow users more possibilities to express their preference QoS values. Iordache and Moldoveanu, 2014 [32] extend the OWL-Q to support users' preferences and trade-offs. They calculate the score vectors of Web services for ranking them. The score vectors are obtained from the comparison of preferences and constraints of a user and Web services.

D'Mello and Ananthanarayana, 2009 [20] extend the OWL-S profile ontology to add QoS properties and business offerings. Their algorithm matches a user's functionality concepts with the functionality ontology; and degrees of match are determined. Web services are ranked based on functional properties, then QoS properties and lastly by business offers values.

In our proposal, Web services are first filtered with the service context. Then, the functional properties of Web services are considered. Finally, the matched functional services are ranked by non-functional properties of Web services. The string matching is the similarity matching by using Synsets of WordNet. In addition, the variation of QoS values over times is considered.

*2.4. Ontology matching*

Ontology matching aims at solving the problem of semantic heterogeneity in information integration and sharing. The goal is to establish correspondences between semantically related entities in different ontologies [33]. The matching of heterogeneous semantic information sources is a hot research topic [22], [34]. We can broadly distinguish three types of ontology matching namely concept-based matching, structure-based matching and instance-based matching. The concept matching, also known as lexical matching, is based on linguistic information. The main idea in using such measures is the fact that usually similar entities have similar descriptions across different ontologies. According to [35], this is the most frequently used ontology mapping method to date. The structure-based matching utilizes structural information in ontology i.e. relationships with entities. It relies on the intuition that elements of two distinct models are similar when their adjacent elements are similar. It leverages the semantics inherited by parents, and passed to children and the structures residing in the ontology graphs to recognize related entities. The instance-based matching relies on the instances which express the semantics of a concept. It is about comparing the instances of the concepts independent of their meta-information [36], [37]. The basic idea of instance-based matching is that the more significant the overlap of common instances of two concepts is, the more related these concepts are. Many ontology matching systems have been developed, sometimes combining several approaches. Nevertheless, according to [36], the instance-based matching remains a promising solution for ontology alignment problems.

In this work, we make use of the instance-based matching. Nevertheless, our goal is not to discover if two instances (or individual) refer to the same real-world entity. The purpose is to retrieve sets of similar instances by comparing them.

3. **Instance-based ontology matching**

Fig. 1 shows the steps towards obtaining sets of ranked Web services corresponding to each business task of a business process. Instances of the business process ontology and of the Web service ontology are generated separately. The business process ontology instance is the result of the transformation of a business process designed with BPMN, according to the BPMN 2.0 specifications. Different methods and tools can be considered for generating the instance of the ontology. For translating XML documents into ontologies, two solutions can be applied. The first one is using XSLT to map two concepts when structures of the source and of the target are known, The second one is using automatic generation by defining generic rules [38], [39]. Available tools are XS2OWL[1], Topbraid[2], OntMalizer[3] framework. The Web services ontology instance is obtained by populating the Web service ontology. Web service descriptions may come from different repository sources as UDDI[4], WSO2[5] Web Services Framework or other service registries.

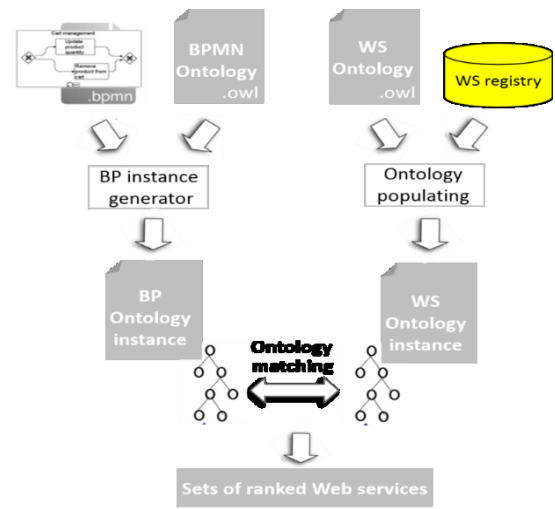

**Fig. 1.** Overview of the process towards selecting Web services for a business process requirement through ontology matching.

---

1 http://www.music.tuc.gr/projects/sw/xs2owl

2 http://www.topquadrant.com/tools/ide-topbraid-composer-maestro-edition

3 https://github.com/srdc/ontmalizer

4 http://www.oracle.com/technetwork/middleware/registry/overview/index.html

5 http://wso2.com

The instance-based ontology matching is performed between the ontology of a business process and the Web service ontology with the multi-criteria selection algorithm. The result comprises sets of Web services ranked according to their score. The score depends on the degree of match found between the properties of the business requirements and the properties of each Web service.

## 4. Business process and web service ontologies

### 4.1. Business process ontology

The business process ontology, called BPOnto, is based on the BPMN 2.0 ontology proposed in [40], which defines the specifications of BPMN 2.0. Before the BPMN 2.0 ontology, two others propositions have been made. The first one is based on the final release of BPMN 1.0 [41]. The second one is based on BPMN 1.1 [42], and it has been adapted to BPMN 1.2. The BPMN 2.0 ontology reflects better the specifications of BPMN 2.0 with a more extensive structure than the other ones. The ontology is written in OWL and composed of two sub-ontologies namely bpmn20base and bpmn20. bpmn20base represents the specifications of the meta-model. bpmn20 extends bpmn20base with the expression of the syntactical requirements taken from the natural text of the BPMN specification. The BPMN 2.0 ontology is composed of 260 classes, 178 object properties and 59 data properties.

### 4.2. Web service ontology

The Web service ontology, called WSOnto, is based on syntactic descriptions of Web services (WSDL files). We extend the syntactic descriptions by introducing QoS attributes in the ontology. To design the ontology, the information comes from different sources: WSDL files, UDDI registries and tracking data of the execution of business processes.

The structure of the WSOnto ontology is presented in Fig. 2. From the version presented in [43], the date-time data property has been added to the Performance concept of Web services and of service's operations. The date-time property allows to manage the performance value that can change over times. Moreover, the QoS are divided into two groups, qualitative and quantitative. Currently, WSOnto has 15 classes, 14 object properties and 16 data properties. In the WSOnto ontology, each service category contains a name, a list of keywords and a set of services. A service is identified by its operations and its QoS properties along with its name, a business name (service provider), a business key (unique identifier of a business entity in UDDI), a service key (unique identifier of a service in UDDI) and its URL (physical location of WSDL file). Operations have two types of properties, functional properties and performance. Functional properties refer to their input and output parameters that are specified by a name and a data type. The performance defines how well an operation was executed. In this current version, WSOnto stores three properties: availability, execution time and number of total calls. At the service level, the QoS is represented by the performance and the security. Performance of a Web service has the same profile as of the operation.

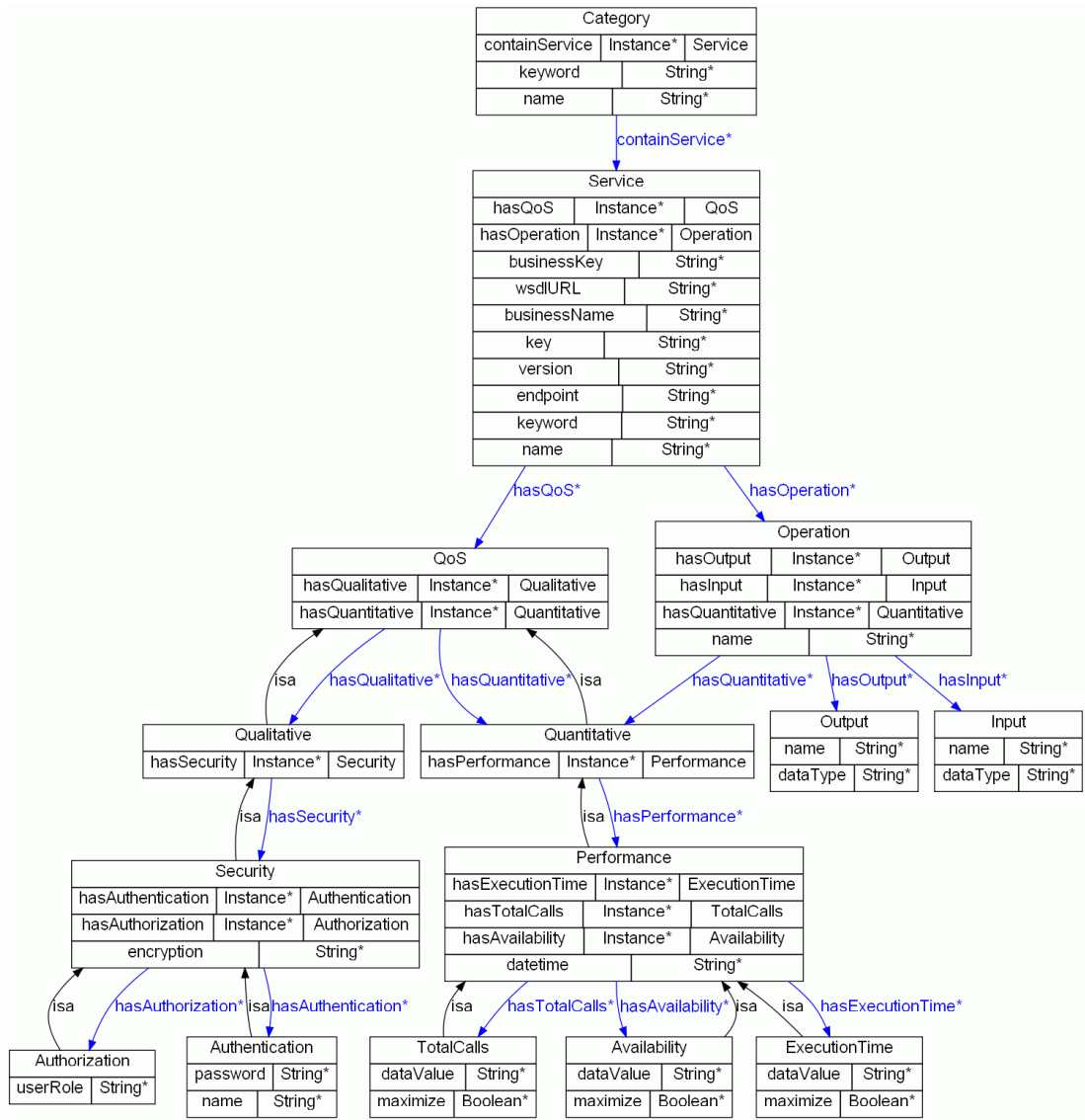

**Fig. 2.** Web service ontology (WSOnto).

## 5. Multi-criteria service selection algorithm

During the matching process, the service selection algorithm looks for correspondences between each business task of a business process defined by a user and Web services stored in the WSOnto ontology. The algorithm is composed of three parts. The first one treats the matching between the context of business tasks in BPOnto ontology and the keywords of service categories in the WSOnto ontology. The second one is related to the functional matching on the inputs and outputs expressed by a user and operations of Web services stored in the WSOnto ontology. Inputs and outputs are specified by string name and string data type. The third one concerns the computation of QoS values. Finally, the algorithm provides a set of ranked Web services in response to each desired business task. The service selection process is illustrated in Fig. 3.

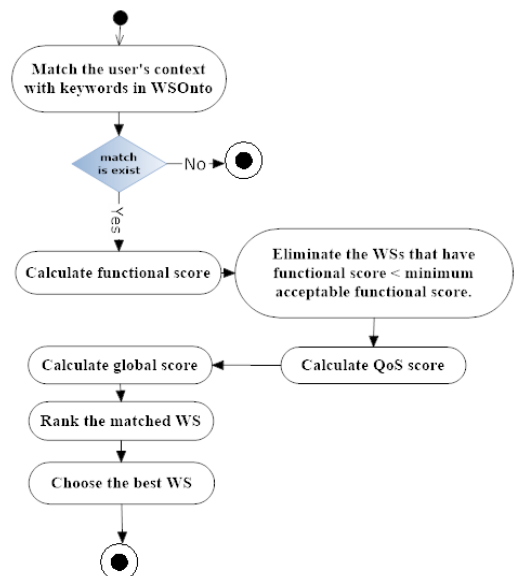

**Fig. 3.** The service selection process.

The objective is to maximize the number of appropriate Web services before calculating the global score that is used to rank Web services. The pseudo code of the service selection algorithm is presented in Fig. 4. The context that is provided by a user is the first filter of the service selection process (Line 5). If the context does not match, then the process stops. Otherwise, the matching process continues by performing the functional matching (Line 10-12). The non-functional (QoS) computation is performed only if the functional score is bigger or equal to the calculated minimum acceptable functional score (Line 14). Finally, Web services are ranked according to their global score (Line 20).

Before the service selection process performs the string matching, all the strings (input and output parameters name, data type and keywords of the context) must be passed through the keyword extraction method introduced in [44] first. This allows reducing the mismatch caused by different naming convenience styles of different users and developers, as well as to solve the synonym problems.

```
1   Traverse BPOnto and WSOnto to obtain a list of business task (BT) and WSs.
2   for each BT
3       // perform context matching
4       for each service category
5           if WSOnto.category.keywords match BPOnto.BT.context != empty
6               SCORE_output = 2*NbOutputUser;
7               for each WS
8                   for each operation
9                       // perform the functional matching
10                      SCORE_FP = functional_matching (WS, OP, BT);
11                      SCORE_min_FP = 2+2*min(NbInputUser, NbInputOperation)+ SCORE_output;
12                      if SCORE_FP >= SCORE_min_FP
13                          // perform the non functional calculation
14                          SCORE_NFP = non_functional_calculation(WS,OP);
15                          global_score = weight*SCORE_FP + (1-weight)*SCORE_NFP;
16                          addWebService (WSSet, WS, OP, global_score);
17                  end for // operation
18              end for // WS
19          end for // service category
20          WSOP = rankWebService(WSSet);
21          addBusTaskList(BTSet, BT, WSOP);
22  end for // business Task
```

**Fig. 4.** Pseudo code of the service selection algorithm.

### 5.1. Context matching

The context of a designed business process is expressed as keywords in the text annotation by a user. A string similarity measure matches this context with the keywords attached to each service' category in the WSOnto ontology. WordNet is integrated in the process. It helps detecting synonyms and managing these problems thanks to its Synset, sets of synonyms. If at least one match is found, the process is considered to be successful. The string matching is done in two steps as follows:

- **Step1**. This step returns true if they are the same string and false otherwise.
- **Step2**. This step relates to WordNet support. It performs when the first step returns false. The algorithm extracts the synonym terms of a user's keywords and the synonym terms of service category's keywords from WordNet. It compares the two sets. If at least one match is found, the context matching is considered to be successful.

This string matching is also applied in other stages of the service selection process when a string matching is required.

### 5.2. Functional matching

A Web service can have many operations. Therefore, the functional matching is done at the operational level. The service selection process calculates the functional score, $SCORE_{FP}$, for all operations of Web services whose context matching succeeded. The functional_matching (WS, OP, BT) function (Fig. 4, line 10), returns $SCORE_{FP}$, the functional score of a Web service's operation. The $SCORE_{FP}$ is obtained from three comparisons.

- **First**: The service selection process compares the numbers of inputs and outputs of a business task (user's request) and the numbers of inputs and outputs of a service's operation. It results in two scores, $SCORE_{nbInput}$ and $SCORE_{nbOutput}$. The $SCORE_{nbInput}$ is obtained from a comparison between the number of inputs of a user's request and of a service's operation, based on the defined method in Fig. 5. The $SCORE_{nbOutput}$ is obtained from a comparison between the number of outputs of a user's request and of a service's operation.
- **Second**: The process performs the string comparison between the names of input and output parameters of a business task (user's request) and candidates Web services' operations. It results into two scores, $SCORE_{strInputName}$ and $SCORE_{strOutputName}$. The $SCORE_{strInputName}$ is obtained from a comparison between the input string names of a user's request and of a service's operation. The $SCORE_{strOutputName}$ is obtained from a comparison between the output string names of a user's request and of a service's operation.
- **Third**: The algorithm performs the string

comparison between the string data types of the input and output parameters of a business task (user's request) and of candidate services' operations. It results into two scores, $SCORE_{strInputDatatype}$ and $SCORE_{strOutputDatatype}$. The $SCORE_{strInputDatatype}$ is obtained from a comparison between the input string data type of a user's request and of a service's operation. The $SCORE_{strOutputDatatype}$ is obtained from a comparison between the output string data types of a user's request and of a service's operation.

Fig. 5 illustrates the score calculation method for a single parameter. The notation "Nb" refers to the number of inputs or outputs. "Equal" means that a user's request and a considered Web service's operation have the same number of inputs (respectively outputs). "Less" means that a user's request has less inputs (respectively outputs) than a considered Web service's operation. "More" means that a user's request has more inputs (respectively outputs) than a considered Web service's operation. The notation "String" refers to the string names of input and output parameters, and of data types.

Regarding the number of parameters, in both cases (inputs and outputs), "Equal" has the highest score. It is considered as the perfect match. Scores of the "Less" and "More" cases of input and output parameters are inverted. Indeed, it is better to have more inputs requested by a user than a service's operation. For output parameters, it considers that the "Less" case is more satisfying. It is better for users to obtain more outputs than requested rather than receiving less outputs. When they receive more, they can filter what they want.

For the name and data type string matching, two situations are considered: equality ("Same") and non-equality ("Different").

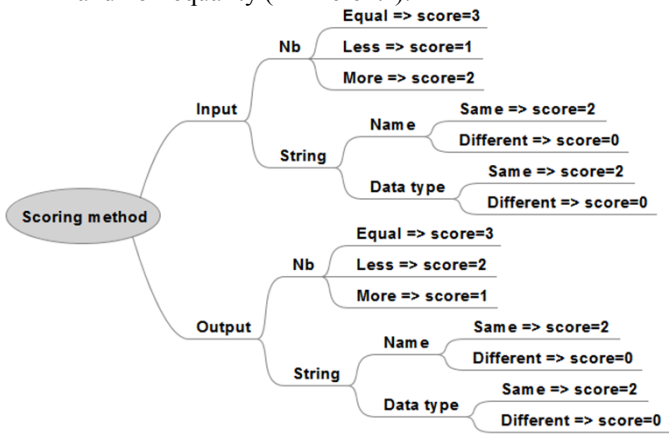

Fig. 5. Scoring method of the functional matching.

The functional score, $SCORE_{FP}$, is calculated by summing the scores of all inputs and outputs based on the scoring method defined in Fig. 5. The $SCORE_{FP}$ is calculated by using equation (1).

$SCORE_{FP} = SCORE_{input} + SCORE_{output}$

$SCORE_{FP} = SCORE_{nbInput} + SCORE_{stringInput}$ $+ SCORE_{nbOuput} + SCORE_{stringOutput}$

$SCORE_{FP} = SCORE_{nbInput} + SCORE_{strInputName}$ $+ SCORE_{strInputDatatype} + SCORE_{nbOutput}$ $+ SCORE_{strOuputName} + SCORE_{strOutputDatatype}$

(1)

To minimize the execution time of the proposed service selection algorithm, some candidate Web services should be removed in this phase. In this case, the minimum acceptable functional score, $SCORE_{min,FP}$, is defined.

A request for Web services can return six possible responses. The input and output parameters below refer to the matched between inputs/outputs of a user's request and a service's operation.

1. One or a collection of Web services that have the same inputs and outputs as requested. It is the best case of all cases.
2. One or a collection of Web services that provide the same outputs, but provide less or more inputs.
3. One or a collection of Web services that provide additional outputs than requested, with the same inputs.
4. One or a collection of Web services that provide additional outputs than requested, but not the same inputs (less or more input parameters).
5. One or a collection of Web services that provide fewer outputs than requested, but with the same inputs.
6. One or a collection of Web services that provide fewer outputs than requested, plus not the same inputs (less or more input parameters).

Case 5 and case 6 do not answer to users' requests. Users need Web services that provide all outputs that they need. For inputs, a service's operation can provide exactly the same parameters, less or more parameters with inputs of a user's request. It is still acceptable because some input parameters might be optional parameters for a service's operation. To avoid obtaining Web services that satisfy only outputs, but not at all inputs, inputs must be as well considered in the calculation of the functional score. In the case 1, 2, 3 and 4 are the cases that satisfy requests of users. In addition, case 4 generates less score comparing to three other cases. Therefore, the minimum acceptable functional score, $SCORE_{min,FP}$, is calculated from the case 4. In the functional matching step of the service selection process, it eliminates all Web services that cannot provide all outputs requested by a user. It considers only Web services that reply to all outputs of a user's request and at least they provide some matched input parameters. The $SCORE_{min,FP}$ is defined in equation (2)

. It is calculated from the sum scores of the matching between what is provided by a user with what is requested by a Web service's operation. It includes the matched score of input numbers, output numbers, inputs' string names, outputs' string names, inputs' string data types and outputs' string data types. The matched score is defined based on the scoring method defined in Fig. 5.

$SCORE_{min,FP} = SCORE'_{nbInput} + SCORE'_{nbOutput} + SCORE'_{strInputName} + SCORE'_{strOutputName} + SCORE'_{strInputDatatype} + SCORE'_{strOutputDatatype}$

Where:

$SCORE'_{nbInput} = 1$, the case when the number of a user's inputs is less than what requested by a service's operation. The proposed service selection algorithm accepts all cases when the number of a user's inputs is equal to, less or more than what requested by a WS's operation.

$SCORE'_{nbOutput} = 2$, the case when the number of a user's outputs is less than what requested by a WS's operation. This proposed algorithm accepts both cases when the number of a user's outputs is less than or equal to what requested by a WS's operation.

$SCORE'_{strInputName} = 2*MIN(NbInput_{user}, NbInput_{operation})$. For one input, if an input string name of a user is the same than what is requested by a WS's operation, then the match score is equal to 2. The $2*MIN(NbInput_{user}, NbInput_{operation})$ signifies that the proposed algorithm accepts only WSs that at least have some match inputs as requested by a user.

$SCORE'_{strOutputName} = 2*NbOutput_{user}$. For one output, if an output string name of a user is the same than an output string name provided by a WS's operation, then the matching score is equal to 2. The $2*NbOutput_{user}$ signifies that the proposed algorithm accepts only WSs that provide the same outputs than requested by users; and when they provide additional output only.

$SCORE'_{strInputDatatype} = SCORE'_{strInputName}$, because they are the score of string matching of inputs. The differences are that one is for the string name and the other is for the string data type. The same case for $SCORE'_{strOutputDatatype} = SCORE'_{strOutputName}$.

Therefore,

$SCORE_{min,FP} = 1 + 2 + 2*2*MIN(NbInput_{user}, NbInput_{operation}) + 2*2*NbOutput_{user}$

$SCORE_{min,FP} = 3 + 4*MIN(NbInput_{user}, NbInput_{operation}) + 4*NbOutput_{user}$  (2)

Web services' operations that have $SCORE_{FP} >= SCORE_{min,FP}$ are kept for further service selection process.

*5.3. QoS calculation*

The Non-Functional Property Score, $SCORE_{NFP}$, is also computed at the operational level like the calculation of the functional property score. The $SCORE_{NFP}$ is calculated only for Web services that validate the context matching, and whose functional property score is at least equal to the score calculated from the equation (2)
. Usually, authors rank Web services by using score calculated from the utility function. The service requester preferences are mapped to values of utility, where higher utility means greater preferences. In order to reflect the influence of the dynamically changing of QoS values over times, the QoS score is calculated in term of the variability of utility scores. Therefore, the $SCORE_{NFP}$ is obtained from the score of the aggregate change ($SCORE_{AC}$) of QoS values and the score of the utility function ($SCORE_{UFt0}$) at the time $t=0$ as expressed in equation (3).

$SCORE_{NFP} = w*SCORE_{UFt0} + (1-w)*SCORE_{AC}$  (3)

With $0 < w < 1$, where w represents the weight.
The $SCORE_{NFP}$ is calculated in four steps, as follows:

**Step1**: Calculate the utility function using equation (4). This utility function is proposed in [45] for calculating the QoS value.

$$UF = \sum_{i=1}^{\alpha} w_i * \left(\frac{q_{ai}(K)-\mu_{ai}}{\sigma_{ai}}\right) + \sum_{j=1}^{\beta} w_j * \left(1 - \frac{q_{bj}(K)-\mu_{bj}}{\sigma_{bj}}\right)$$  (4)

Where:
The QoS attributes are divided into two categories, maximize and minimize attributes. The maximize attributes are the attributes whose values need to be maximized. The minimize attributes are attributes whose values require to be minimized.
$q_{ai}$: refers to maximized QoS attributes (the higher the value, the better the quality). For examples, reliability, availability, etc.
$q_{bj}$: refers to minimize QoS attributes (the higher the value, the lower the quality). Some examples of these attributes are price, response time, etc.
α: number of the maximized QoS attributes
β: number of the minimized QoS attributes
w: weight of QoS attributes ($0 < w_i, w_j < 1$) and $\sum_{i=1}^{\alpha} w_i + \sum_{j=1}^{\beta} w_j = 1$
$\mu_{ai}$: average value of all $q_{ai}$ attributes
$\mu_{bi}$: average value of all $q_{bj}$ attributes
$\sigma_{ai}$: standard deviation of all $q_{ai}$ attributes
$\sigma_{bi}$: standard deviation of $q_{bj}$ attributes

**Step 2**: Calculate the change (C) of QoS values of a Web service that changes over n time gaps ($t_0$ to $t_{n-1}$). The change value allows considering the variation of QoS values of a Web service's operation over times. The formula to calculate the change value is defined in equation (5). The time here refers to the time when the process for calculating QoS values was run. The granularity of the time gap can be a week or a month depending on a company's choice. Web services can be created at different times. Therefore, to compare between two Web services, only the last n numbers of time gaps is considered when calculating the changing values of QoS.

$$C_{ti,j} = \left(\frac{UF_j}{UF_i} - 1\right) \text{ with } i < j$$  (5)

Where:

$UF_j$: value of the utility function at time $t_j$
$UF_i$: value of the utility function at time $t_i$
**Step 3**: Calculate the $SCORE_{AC}$ value of each Web service's operation over the last n numbers of time gaps using equation (6).

$$SCORE_{AC} = C_{t0,1} + C_{t1,2} + \cdots + C_{tn-2,n-1} \quad (6)$$

**Step 4**: Calculate the non-functional QoS score ($SCORE_{NFP}$) using equation (3). The $SCORE_{UFt0}$ is the normalized score of UF value at time $t_0$ based on Fig. 6. The normalization is done to solve the small value of UF score (range of [0..1]) comparing to the $SCORE_{AC}$. The w value is chosen by users in order to ponder the effect of the utility function variation. When $w=1$, fluctuations of the utility function over times are not taken into account.

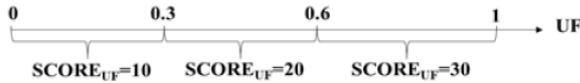

**Fig. 6.** Normalization score solution of utility function score.

*5.4. Service selection algorithm*

The service selection algorithm ranks the matched functional services by using a global score. This global score is the sum of the multiplication of weights with a functional score ($SCORE_{FP}$) and a non-functional score ($SCORE_{NFP}$). Users define the importance of a functional score ($SCORE_{FP}$) compare to a non-functional score ($SCORE_{NFP}$) by defining weights. The global score can be calculated using equation (7).

$$global\_score = w*SCORE_{FP} + (1-w)*SCORE_{NFP} \quad (7)$$

**6. Implementation**

The proposed service selection algorithm that performs the instance-based matching is implemented within a framework called BPMNSemAuto [47]. BPMNSemAuto supports the entire process from the design of a business process until its implementation. It takes as input the designed business process with text-annotations that describe the functionality of each business task. The weights of the QoS attributes, the functional properties and context of the business tasks are also provided by the user. It finally generates an executable business process as output. In its actual version, the framework embodies four modules. The "Existing SOA Infrastructure" module represents the pool of the available Web services published by providers in UDDI registries. The Web services are grabbed from registries and stored in the WSOnto ontology. Fig. 7 shows a list of individuals of WSOnto with two Web services, "authentication" and "sendEmail". The "authentication" Web service is identified by an instance named ws.15.09.2013.08.43.40. The "sendEmail" Web service is identified by an instance named ws.15.09.2013.09.43.45. The format is ws.Date.Time. ws stands for web service and Date.Time is the date when the Web service has been created in the ontology. Each Web service has two operations identified by instances whose name is of the form opX.wsY, where *X* is the operation number followed by the name of the Web service instance to which the operation belongs to. Operations op1 of each Web service have two inputs and one output parameters. Parameter instances name are of the form inputZopXwsY and outputZopXwsY. Z is the input/output number, X is the operation number followed by the name of the Web service instance to which the operation belongs to. Operation QoS attributes (availability, number of calls, execution time) appear as attributeNamepfDateopXwsY. attributeName is the name of the attribute (aval, call, exec), **pf** stands for performance, Date is the date at which the script for updating the QoS values has been run. *Date* is followed by the operation and by the Web service instance to which the operation belongs to. Two series of QoS values are available. They correspond to two different dates (14-01-2014 and 14-02-2014). Our system works on a monthly based time gap. Note that we do not have any values for op2 QoS attributes. QoS values are available only when operations have been executed in a business process. Lines starting by pf are related to the performance object of a Web service and to the performance object of an operation.

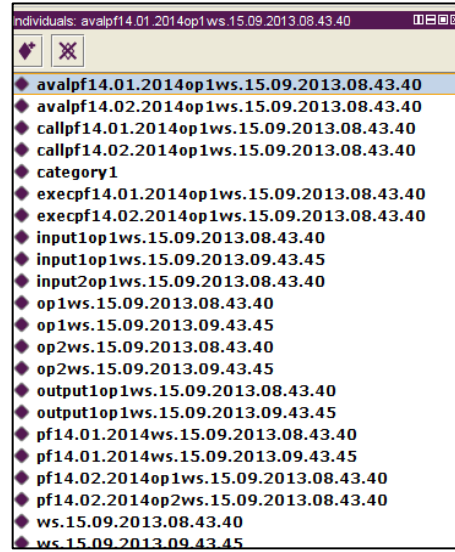

**Fig. 7.** Extract of WSOnto with two Web services (ws), operations (op), parameters (input, output), Web service category, QoS attributes (avalpf, callpf, execpf) (View in Protégé).

Fig. 8 shows the content of the "authentication" Web service. Its three keywords are listed (system, authentication, login). Its two operations are identified respectively by op1ws.15.09.2013.08.43.40 and

op2ws.15.09.2013.08.43.40. The location of the WSDL file is http://159.84.79.144:9763/services/Authentication?wsdl. Note that the type of the data properties is mentioned.

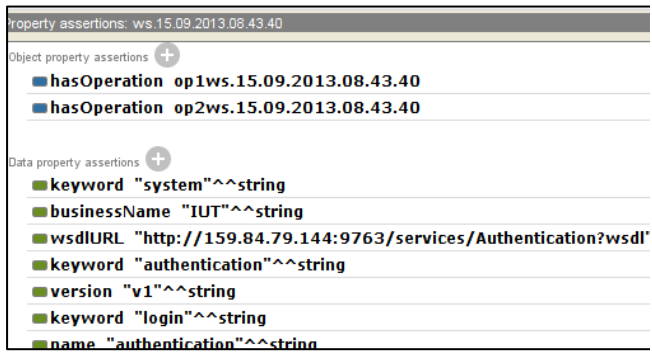

**Fig. 8.** Details of a Web service in WSOnto: Web service name, operations, version, keywords and location of WSDL file (View in Protégé).

Fig. 9 shows the information of the "op1ws.15.09.2013.08.43.40" individual. The hierarchy specifies the type of this individual (operation), its name (login), its parameters (hasInput, hasOuptut) and its global QoS.

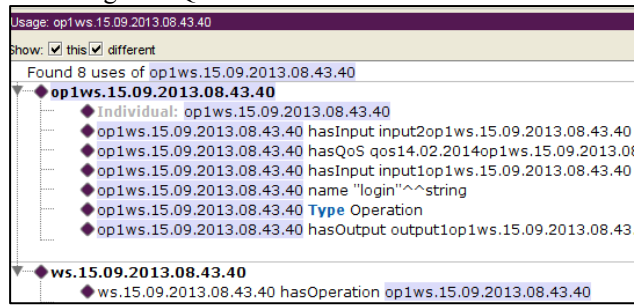

**Fig. 9.** Information of a Web service' operation.

The "Semantic Representation of Users Requirements" module outputs BPOnto ontology, the business process ontology. A business process is designed by using any support business process modeling tools such as Activiti[1] or JDeveloper[2]. For each business task, users provide a textual description in terms of inputs, outputs and context. The input and output parameters are represented by the couple (variable-name, variable-value). The context is defined by a set of keywords that describe the business task. Fig. 10 illustrates a business process of sending an email, and the description of each task. The first task is the authentication task with two input parameters (username and password), one output parameter (authentication), and a context defined by the keywords, "authentication" and "login". The second task is the sending of the email. It contains three input parameters (address of the sender, the address of the receiver and the content of the email), and one output parameter "reply" which is a boolean value expressing the state of the message i.e. sent successfully or not.

An XML file describes the designed business process and its business tasks. At this step, the file is enriched by the user, through a "User Interface", with requirements on QoS. Those requirements are the weights for each QoS property and weights of scores of the functional and non-functional properties of Web services. From the XML file, the semantic representation of the business process is generated. This task is performed by a "Semantic Transformer" using an existing XSLT transformation. The transformer generates the BPOnto ontology in turtle format, based on the BPMN 2.0 ontology. An ontology reasoner allows detecting whether the designed business process diagram is constructed with respect to the specifications of BPMN 2.0 or not. In the current version of the framework, Pellet, FaCT++ or Hermit reasoners can be used [46], [47].

Fig. 11 illustrates the individuals/instances of the BPOnto ontology for the example defined in Fig. 10. Instances of service task (servicetask1, servicetask2) represent business tasks. Instances of text-annotation (textannotation1, textannotation2) represent the requirement expression of each business task of the user. Instances of association (association1, association2) define the relationship between a service task and a text-annotation. Instances of sequence-flow (flow1, flow2, flow3) define the interaction between an interaction node to another interaction node. For example the flow1 defines the sequence flow from startevent1 to servicetask1. In our service selection algorithm, only instances of service task, text-annotation and association are considered as important to find the most appropriate Web services to implement business tasks.

---

[1] http://activiti.org
[2] http://www.oracle.com/technetwork/developer-tools/jdev/overview/index.html

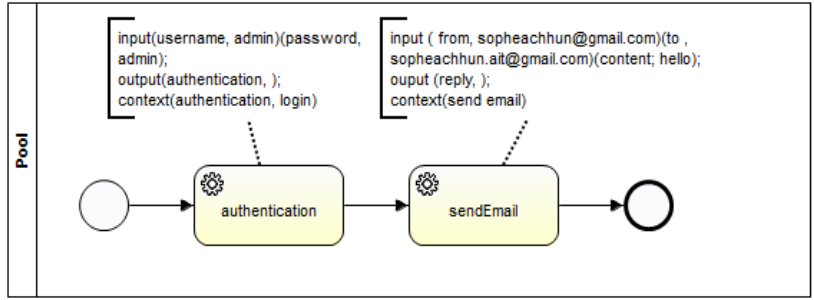

**Fig. 10.** A business process that contains two tasks with attached information (input and output parameters, context).

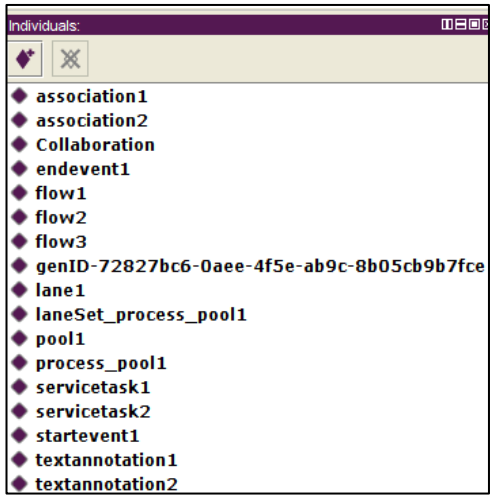

**Fig. 11.** Instances of BPOnto ontology related to the business process defined in Fig. 10.

Fig. 12 shows the content of text annotation2 instance. Its identity is followed by text content (input, output and context).

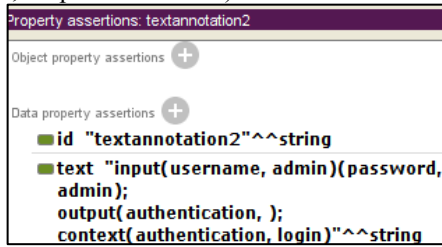

**Fig. 12.** Content of textannotation2 of BPOnto ontology.

Fig. 13 shows the content of association2. We can see that servicetask1 (authentication) is associated with textannotation2.

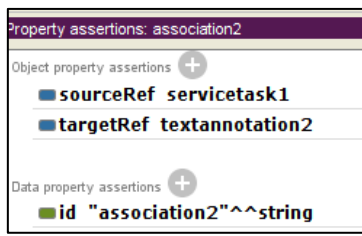

**Fig. 13.** Content of association2 of BPOnto ontology.

In the "Implementation of Business Application" module, the "Semantic Matching Engine" performs the matching between BPOnto and WSOnto ontologies. This module finally generates the executable business process. Fig. 14 shows the ontology matching process that looks for correspondences between instances of the two ontologies, in order to find appropriate Web services to implement business tasks.

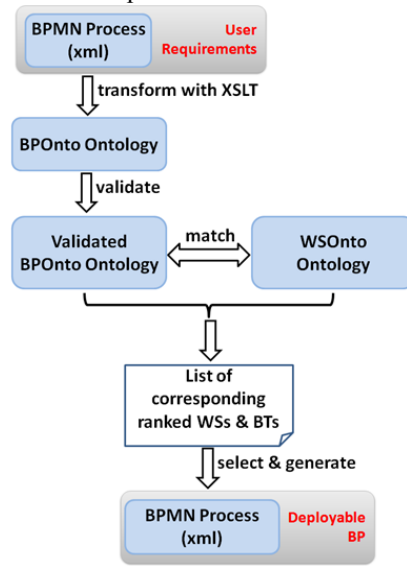

**Fig. 14.** Implementation steps of business processes.

## 7. Conclusion and future work

In this paper, we propose a solution for realizing business processes by leveraging the Web service paradigm. Our proposition makes use of ontologies to represent business processes as well as sets of available Web services. These choices allow coping with two major challenges faced by companies: the agility of applications to manage ongoing changes and the automation of business processes.

Business processes are designed using the well-known standard modeling notation, BPMN2.0. A transformation of the designed business processes results in an ontology expressing users' needs. The correctness of business process diagrams is checked with the help of an ontology reasoner.

The BPMNSemAuto framework implements a service selection algorithm based on a semantic matching between ontologies. This algorithm is based on three criteria: the context, functional and non-functional properties. It performs the matching between the context expressed by a user and the context of service categories, the matching between the number of inputs and outputs, and string comparison between name and data-type of parameters. It calculates the value of three QoS attributes for each Web service. The originality of the calculation of the QoS attributes values relies on the fact that we take into account the variation of the values over a number of time gaps.

The context is determinant to continue or not the selection process. If the matching goes on, a score is provided to Web services by the functional matching and QoS calculating. Candidate Web services are ranked according to the scores, to the weights on the QoS attributes, to functional and non-functional weights provided by users.

The richness of BPMN and the power of ontologies coupled with a sophisticated selection algorithm is a promising solution to reach dynamic and automatic business process implementation. The service selection is one salient functionality of BPMNSemAuto. Nevertheless, the algorithm only implements the selection of atomic Web services for business tasks. We are now developing a composition algorithm. Note that WSOnto has been designed for syntactic descriptions of Web services. It is populated with WSDL file information. The ontology can be easily extended to take into account semantic descriptions of Web services.

**Acknowledgement**

This project has been funded with support from the European Commission (EMA2-2010- 2359 Project). This publication reflects the views only of the author, and the Commission cannot be held responsible for any use which may be made of the information contained therein.**References**

[1] B. P. Model, "Business Process Model and Notation (BPMN) Version 2.0," *Object Manag. Group Specif.*, 2011.

[2] N. Lohmann, E. Verbeek, and R. Dijkman, "Petri net transformations for business processes–a survey," in *Transactions on Petri Nets and Other Models of Concurrency II*, Springer., Springer, 2009, pp. 46–63.

[3] D. Georgakopoulos, M. Hornick, and A. Sheth, "An overview of workflow management: From process modeling to workflow automation infrastructure," *Distrib. Parallel Databases*, vol. 3, no. 2, pp. 119–153, 1995.

[4] O. M. G. Specification, "OMG Unified Modeling Language (OMG UML), Superstructure, V2.1.2," *Object Manag. Group*, 2007.

[5] A. Arkin and others, "Business process modeling language," *BPMI Org*, pp. 1–96, 2003.

[6] D. Krafzig, K. Banke, and D. Slama, *Enterprise SOA: service-oriented architecture best practices*. Prentice Hall Professional, Chapter. 4, 2005.

[7] D. Mukhopadhyay and A. Chougule, "A survey on web service discovery approaches," in *Advances in Computer Science, Engineering & Applications*, Springer, 2012, pp. 1001–1012.

[8] A. Pradnya Khutade and B. Rashmi Phalnikar, "QOS BASED WEB SERVICE DISCOVERY USING OO CONCEPTS," *Int. J. Adv. Technol. Eng. Res. IJATER*, vol. 2, no. 6, pp. 81–86, 2012.

[9] S. Chaari, Y. Badr, F. Biennier, C. BenAmar, and J. Favrel, "Framework for web service selection based on non-functional properties," *Int. J. Web Serv. Pract.*, vol. 3, no. 2, pp. 94–109, 2008.

[10] Y. Wu and X. Wang, "Applying Multi-objective Genetic Algorithms to QoS-aware Web Service Global Selection.," *Adv. Inf. Sci. Serv. Sci.*, vol. 3, no. 11, pp. 1–10, 2011.

[11] S. S. Yau and Y. Yin, "QoS-Based Service Ranking and Selection for Service-Based Systems," 2011, pp. 56–63.

[12] Z. Junfeng, "Web Service Selection Based on the Interval Grey Number of QoS," *Int. J. Adv. Comput. Technol.*, vol. 5, no. 5, pp. 40–47, Mar. 2013.

[13] R. Lu and S. Sadiq, "A survey of comparative business process modeling approaches," in *Business Information Systems*, 2007, pp. 82–94.

[14] A. Correia and F. B. e Abreu, "Adding Preciseness to BPMN Models," *Procedia Technol.*, vol. 5, pp. 407–417, Jan. 2012.

[15] R. Chinnici, J.-J. Moreau, A. Ryman, and S. Weerawarana, "Web services description language (wsdl) version 2.0 part 1: Core language," *W3C Recomm.*, vol. 26, p. 19, 2007.

[16] D. Martin, M. Burstein, J. Hobbs, O. Lassila, D. McDermott, S. McIlraith, S. Narayanan, M. Paolucci, B. Parsia, T. Payne, and others, "OWL-S: Semantic markup for web services," *W3C Memb. Submiss.*, vol. 22, pp. 1–30, 2004.

[17] J. De Bruijn, C. Bussler, J. Domingue, D. Fensel, M. Hepp, M. Kifer, B. König-Ries, J. Kopecky, R. Lara, E. Oren, and others, "Web service modeling ontology (wsmo)," *Interface*, vol. 5, pp. 1–33, 2005.

[18] J. Kopecky, T. Vitvar, C. Bournez, and J. Farrell, "Sawsdl: Semantic annotations for wsdl and xml schema," *Internet Comput. IEEE*, vol. 11, no. 6, pp. 60–67, 2007.

[19] R. Akkiraju, J. Farrell, J. A. Miller, M. Nagarajan, A. Sheth, and K. Verma, "Web service semantics-wsdl-s," pp. 1–35, 2005.

[20] D. A. D'Mello and V. S. Ananthanarayana, "Semantic Web Service Selection Based on Service Provider's Business Offerings," *IJSSST*, vol. 10, no. 2, pp. 25–37, Mar. 2009.

[21] A. D'Ambrogio, "A WSDL extension for performance-enabled description of web services," in *Computer and Information Sciences-ISCIS 2005*, Springer, 2005, pp. 371–381.

[22] P. Shvaiko and J. Euzenat, "Ontology matching: state of the art and future challenges," *Knowl. Data Eng. IEEE Trans. On*, vol. 25, no. 1, pp. 158–176, 2013.

[23] L. Predoiu, C. Feier, F. Scharffe, J. de Bruijn, F. Martin-Recuerda, D. Manov, and M. Ehrig, "D4.2.2 State-of-the-art survey on Ontology Merging and Aligning V2," *EU-IST Integr. Proj. IST-2003-506826 SEKT*, 2005.